\documentclass[5p,authoryear,review,number,sort&compress]{elsarticle}

\usepackage{amssymb}
\usepackage[hidelinks,colorlinks]{hyperref}
\graphicspath{{figs_design/}{figs_prelim/}{figs_montecarlo/}{figs_peakfinding/}{figs_performance/}}

\newcommand{\degree}{\ensuremath{^\circ}}

\newcommand{\figref}[1]{\hyperref[fig:#1]{Figure~\ref*{fig:#1}}}
\newcommand{\figrefii}[2]{Figures~\ref{fig:#1} and~\ref{fig:#2}}

\newcommand{\secref}[1]{\hyperref[sec:#1]{Section~\ref*{sec:#1}}}

\newcommand{\appref}[1]{\hyperref[app:#1]{\ref*{app:#1}}}

\newcommand{\eqnref}[1]{\hyperref[eqn:#1]{Equation~\ref*{eqn:#1}}}

\newcommand{\tabref}[1]{\hyperref[tab:#1]{Table~\ref*{tab:#1}}}

\title{The SKA Particle Array Prototype: The First Particle Detector at the Murchison Radio-astronomy Observatory}

\author[manchester]{J.D.~Bray\corref{corresponding}}
\author[icrarcurtin]{A.~Williamson}
\author[icrarcurtin]{J.~Schelfhout}
\author[icrarcurtin]{C.W.~James\corref{corresponding}}
\author[manchester]{R.E.~Spencer}
\author[icraruwa]{H.~Chen}
\author[manchester]{B.D.~Cropper}
\author[curtin]{D.~Emrich}
\author[StAndrews]{K.M.L.~Gould}
\author[kit]{A.~Haungs}
\author[manchester]{W.~Hodder}
\author[manchester]{T.~Howland}
\author[kit,vub]{T.~Huege}
\author[icrarcurtin]{D.~Kenney}
\author[curtin]{A.~McPhail}
\author[manchester]{S.~Mitchell}
\author[manchester]{I.C. Ni\c{t}u}
\author[csiro]{P.~Roberts}
\author[manchester]{R.~Tawn}
\author[icrarcurtin]{J.~Tickner}
\author[icrarcurtin]{S.J.~Tingay}

\cortext[corresponding]{Corresponding authors: \texttt{justin.bray@manchester.ac.uk} \texttt{clancy.james@curtin.edu.au}}

\address[manchester]{JBCA, Dept.\ of Physics \& Astronomy, University of Manchester, Manchester M13 9PL, UK}
\address[icrarcurtin]{International Centre for Radio Astronomy Research, Curtin University, Bentley, WA 6102, Australia}
\address[icraruwa]{International Centre for Radio Astronomy Research, University of Western Australia, Crawley 6009, Australia}
\address[curtin]{Curtin Institute of Radio Astronomy, Curtin University, GPO Box U1987, Perth WA 6845}
\address[StAndrews]{School of Physics and Astronomy, University of St. Andrews, North Haugh, St. Andrews KY16 9SS, UK}
\address[kit]{Institut f\"{u}r Kernphysik (IKP), Karlsruhe Institute of Technology, P.O.\ Box 3640, Karlsruhe 76021,Germany}
\address[vub]{Astrophysical Institute, Vrije Universiteit Brussel, Pleinlaan 2, 1050 Brussels, Belgium}
\address[csiro]{CSIRO Astronomy \& Space Science, Epping, NSW 1710, Australia}

\begin{document}

\begin{abstract}
We report on the design, deployment, and first results from a scintillation detector deployed at the Murchison Radio-astronomy Observatory (MRO). The detector is a prototype for a larger array --- the Square Kilometre Array Particle Array (SKAPA) --- planned to allow the radio-detection of cosmic rays  with the Murchison Widefield Array and the low-frequency component of the Square Kilometre Array. The prototype design has been driven by stringent limits on radio emissions at the MRO, and to ensure survivability in a desert environment.
Using data taken from Nov.\ 2018 to Feb.\ 2019, we characterize the detector response while accounting for the effects of temperature fluctuations, and calibrate the sensitivity of the prototype detector to through-going muons. This verifies the feasibility of cosmic ray detection at the MRO. We then estimate the required parameters of a planned array of eight such detectors to be used to trigger radio observations by the Murchison Widefield Array.
\end{abstract}

\begin{keyword} 
 Cosmic rays \sep
 Scintillation detectors \sep
 Murchison Radio-astronomy Observatory \sep
 Murchison Widefield Array \sep
 Square Kilometre Array
\end{keyword}

\maketitle

\section{Introduction}
\label{sec:intro}

Cosmic rays are energetic particles --- mostly protons and atomic nuclei --- from outside the Solar System. At very high energies, cosmic rays impacting the upper atmosphere generate `extensive air showers' (EAS) of secondary particles, some of which are detectable at ground level. These EAS also generate sub-microsecond bursts of radio waves, which can be studied by radio telescopes \citep{huege2016}. The ground pattern of radio emission reflects the charge distribution within EAS, which in turn contains information on the primary particle itself, and the physics of its high-energy collision.  The LOFAR radio telescope has demonstrated that measuring the ground pattern of $\sim10^{17}$\,eV cosmic rays with a dense array of radio receivers yields a precise reconstruction of EAS properties \citep{schellart2013,buitink2014}, and new information about cosmic-ray composition \citep{buitink2016}, while the Auger Engineering Radio Array (AERA) project at the Pierre Auger Observatory has shown that radio detection is an accurate energy estimator \citep{aab2016,aab2016c}.

Radio observations of EAS are used by several astroparticle physics experiments, such as Tunka-Rex \citep{bezyazeekov2018} and ANITA \citep{gorham2009}. The success of cosmic ray detection with LOFAR, i.e.\ with an astronomical telescope, has motivated a similar experiment at OVRO-LWA \citep{monroe2019}. The Square Kilometre Array's (SKA's) High Energy Cosmic Particles Focus Group has suggested using the SKA's low-frequency telescope, SKA1-low, to study radio emission from cosmic rays. The dense core and wide bandwidth (300\,MHz) of this instrument promises to resolve cosmic ray EAS with `ultimate precision' \citep{huege2014}.

SKA1-low is to be deployed at the Murchison Radio-astronomy Observatory (MRO), a remote radio-quiet site in outback Western Australia.  It is home to several radio telescopes, in particular, the SKA1-low precursor instrument, the Murchison Widefield Array (MWA) \citep{tingay2012,wayth2018,beardsley2019}. This is an aperture-array radio telescope consisting of 256 tiles of 16 dipoles each. It is capable of observing over the range 70--300\,MHz with  bandwidth of 30.72\,MHz. In compact configuration, data are returned from over 100 tiles in the central 300\,m region, comparable in size to the `superterp' used to study cosmic rays at LOFAR. As such, the MWA is an ideal instrument with which to test technologies for detecting cosmic rays with the SKA, and explore the science enabled by observing cosmic rays at higher frequencies than LOFAR.

To use the MWA, and ultimately the SKA, for studying cosmic rays requires analyzing the voltages from each spatial element individually at inverse-bandwidth time resolution, i.e.\ with no time-averaging or frequency decomposition. As cosmic ray events are rare (approximately one cosmic ray above $10^{17}$\,eV per hour over the MWA core, using the flux of \citet{apel2012}), a method is required to identify cosmic ray events and trigger the analysis of radio data. LOFAR has successfully used a small particle detector array, LORA \citep{thoudam2011}, for this purpose, and similar arrays are planned for experiments with the MWA and, later, SKA1-low. This is the SKA Particle Array (SKAPA) project, and this paper describes the deployment of the first prototype SKAPA particle detector at the MRO.

This article is structured as follows. \secref{design} describes the design of the prototype SKAPA system, while \secref{prelim} contains a characterization of its performance from its output data. \secref{montecarlo} outlines a simulation analysis of the behavior of the SKAPA particle detector from which we derive, in \secref{performance}, its expected sensitivity to high-energy particles.  Finally, in \secref{conclusion}, we discuss the implications of this analysis for the future development and deployment of the full SKAPA system.

\section{System Design}
\label{sec:design}

\begin{figure}
    \centering
    \includegraphics[width=\columnwidth]{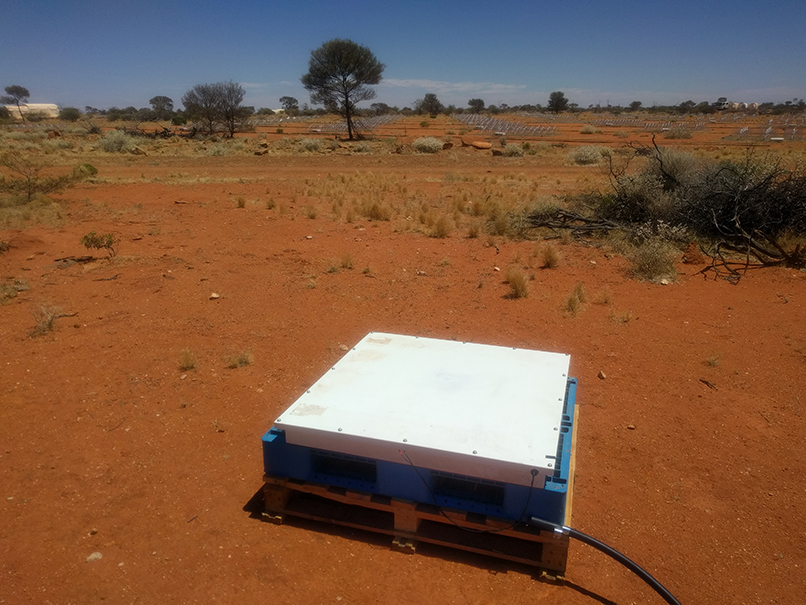}
    \caption{The SKAPA particle detector deployed at the MRO: a \mbox{$1.1\times1.1$}~m$^2$, 50~kg prototype for a future air-shower array.  In the background are tiles of the south hex of the MWA (see \figref{mro_map}).  The power and data cables (see \figref{system}) are routed through the pipe seen in the foreground.  The detector is elevated to mitigate the risk of flooding damage.}
    \label{fig:deployed}
\end{figure}

\begin{figure}
    \centering
    \includegraphics[width=\columnwidth]{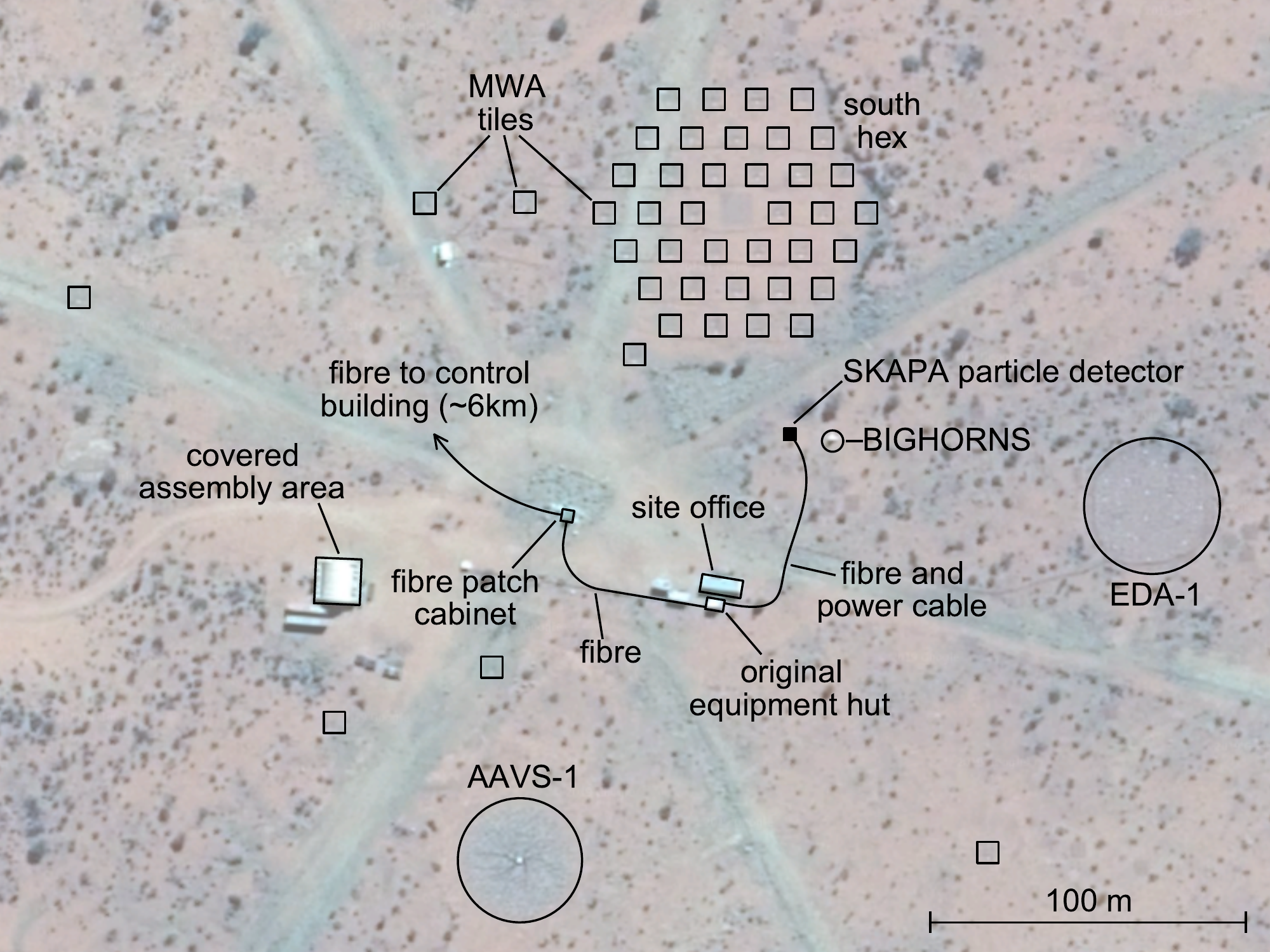}
    \caption{Layout of the SKAPA particle-detector deployment and surrounds.  Power and data connections for the particle detector run to the original equipment hut, fiber patch cabinet and control building, as shown in \figref{system}.  The particle detector is deployed between several radio instruments including tiles of the MWA, mostly in its south hex~\citep{tingay2012}; the first MWA Engineering and Design Array \citep[EDA-1;][]{wayth2017}; a BIGHORNS antenna~\citep{sokolowski2015}; and a station of \mbox{AAVS-1}~\citep{hall2016}, an SKA technology demonstrator.  All of these are sufficiently close (\mbox{$\lesssim 200$}\,m) to detect cosmic-ray events in coincidence with the SKAPA prototype.  Imagery \textcopyright{} 2019 CNES/Airbus/Google.}
    \label{fig:mro_map}
\end{figure}

The deployed SKAPA prototype system consists of a single particle detector (see \figref{deployed}) and its associated support systems, providing electrical power and processing the detector signal output.  The deployment makes heavy use of existing infrastructure at the MRO site, exploiting in-place fiber, networking and mains power.  The only new cabling deployed for this project is the final 100~m run of fiber and power cable to the particle detector from the original equipment hut, which is a small, RF-shielded, air-conditioned enclosure (see \figref{mro_map}).

The particle detector is located adjacent to the south hex of the MWA, one of three dense clusters of antennas that constitute the MWA core.  The close proximity to the core of the MWA allows for the possibility of testing for coincidence with radio emission from air showers, and also enables future iterations of this detector to be used for triggering the MWA to measure any radio emission from these air showers.  There are several nearby radio instruments that may also be used for such future tests: in particular, AAVS-1 will be critical for developing the capability to work with SKA1-LOW, for which it is a prototype.

\begin{figure}
    \centering
    \includegraphics[width=\columnwidth]{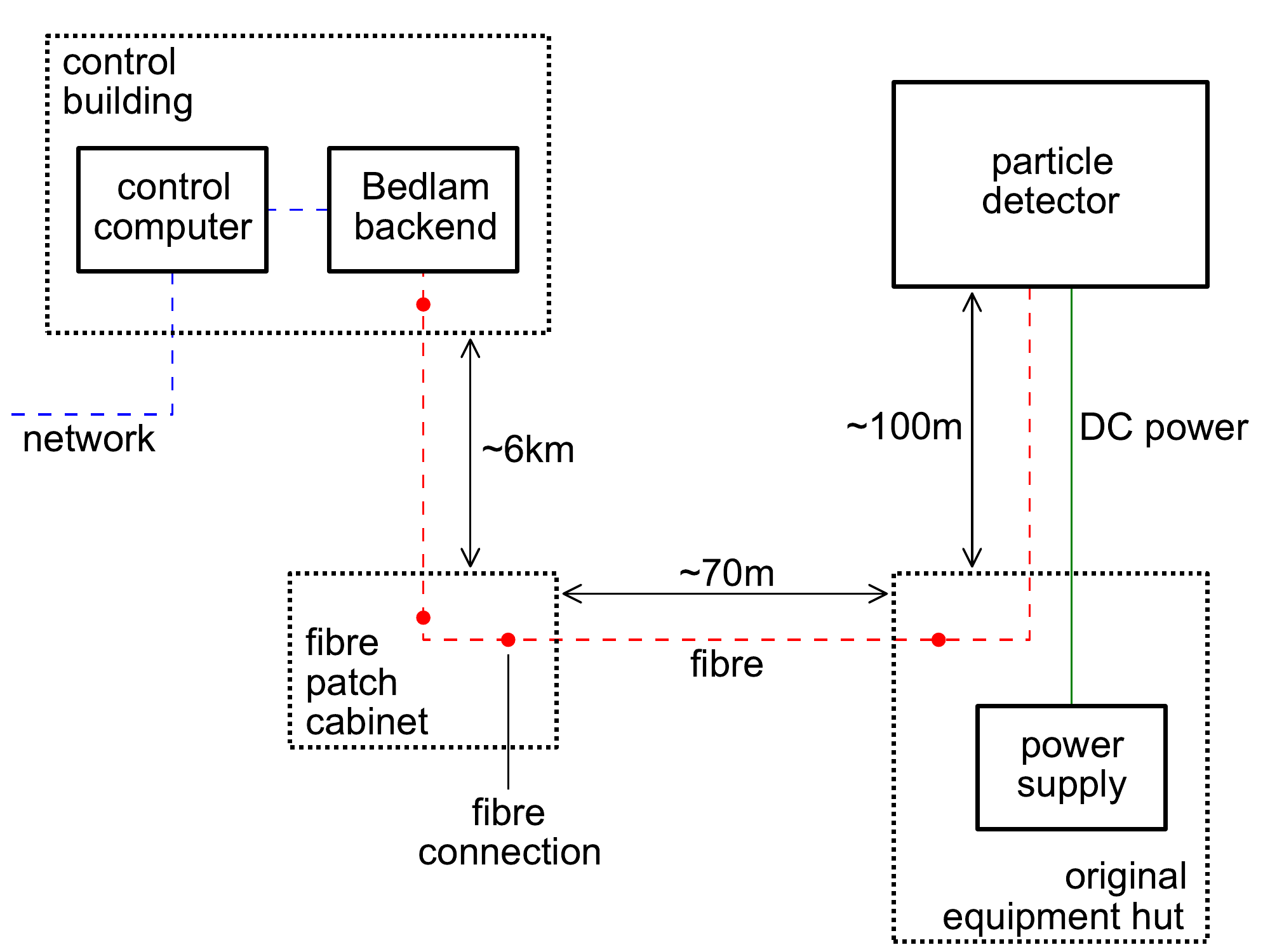}
    \caption{Conceptual overview of the deployed SKAPA prototype system (cf.\ \figref{mro_map}).  DC power for the particle detector is provided from the original equipment hut, while fiber for data return is patched through the original equipment hut and a fiber patch cabinet, eventually reaching the Bedlam backend in the MRO control building around 6\,km away.}
    \label{fig:system}
\end{figure}

\subsection{Signal path}
\label{sec:sigpath}

\begin{figure}
    \centering
    \includegraphics[width=\columnwidth]{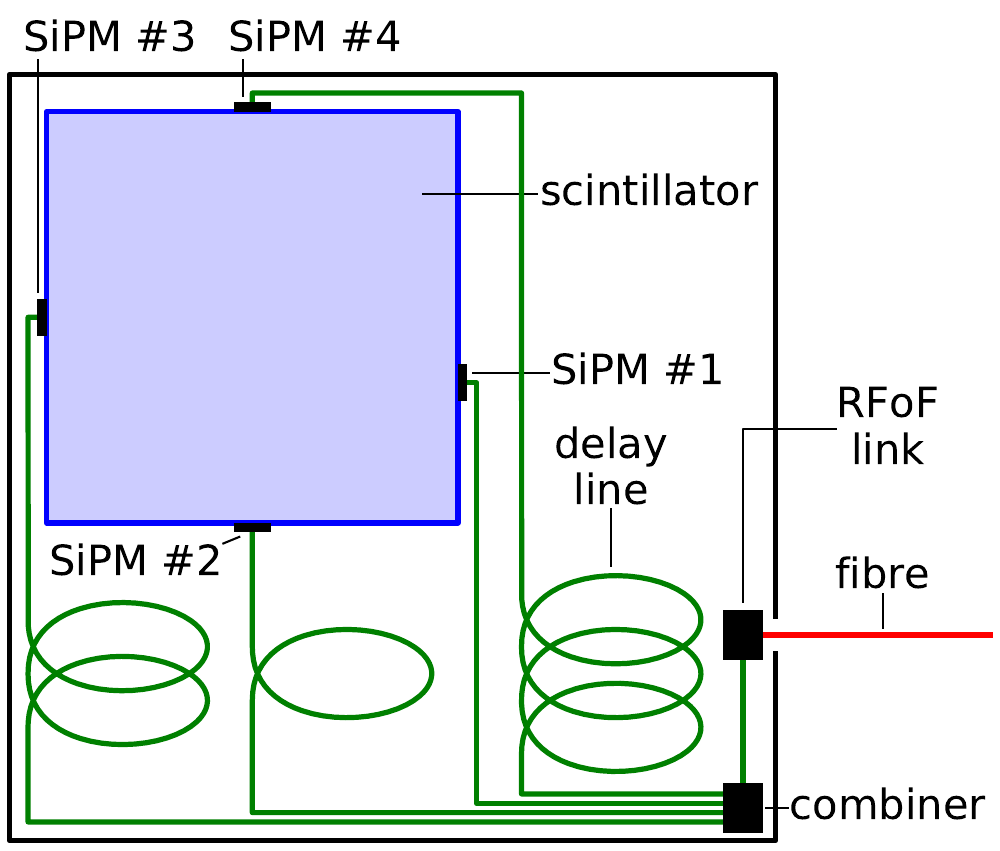}
    \caption{Simplified internal layout of the particle detector (not to scale).  A light pulse from a high-energy particle interacting in the scintillator panel can be detected simultaneously by the four SiPMs.  After passing through delay lines, these are combined, so the pulses detected on SiPMs~\#1--4 will appear in sequence on the output.}
    \label{fig:schematic}
\end{figure}

The key part of the detector is a $90~{\rm cm} \times 90~{\rm cm} \times 3~{\rm cm}$ panel of scintillator material contained within a padded aluminium box (see \figref{schematic}). The scintillator panel is viewed by four silicon photomultipliers (SiPMs), located on each of the four thin sides of the panel, and affixed with optical gel. As a charged particle moves through the scintillator, a short pulse of light is emitted that is detected by these SiPMs. The output produced is a short pulse with power approximately proportional to the number of photons detected.

\begin{figure}
    \centering
    \includegraphics[width=\columnwidth]{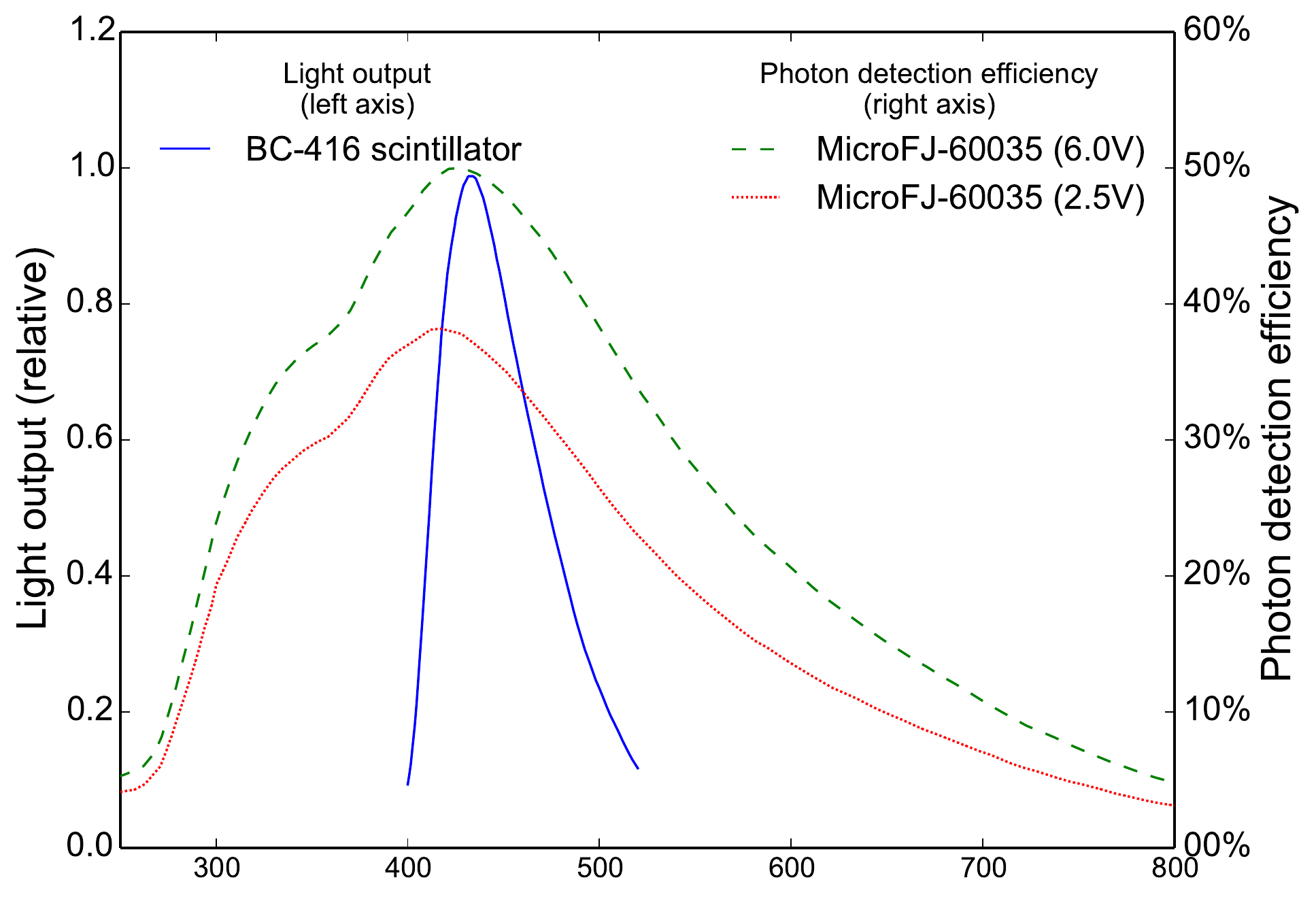}
    \caption{The emission spectrum of the BC-416 scintillator used in this experiment~\citep{kriegleder1992}, compared with the photon detection efficiency of the SensL MicroFJ-60035 SiPMs when operated at 2.5V or 6.0V over their breakdown voltage~\citep{sipmdata}.  The scintillator emission peaks at a wavelength of 430~nm, at which the SiPMs are near their peak efficiency.}
    \label{fig:spect}
\end{figure}

The SiPMs used for this experiment are SensL J-Series 60035 low-light sensors~\citep{sipmdata}, which are designed specifically for robustness, low operating voltage and good temperature stability. Each sensor is located on one of the four edges of the scintillator panel and consists of four smaller SiPM panels, each with an area of $6\times6$~mm$^2$ and containing $22\,292$ microcells, connected in parallel. The SiPMs have a photon detection efficiency (PDE) that peaks at 430 nm (\figref{spect}) and changes with the supplied overvoltage (the difference between the bias and breakdown voltage).

The scintillator itself is a remnant of the KArlsruhe Shower Core and Array DEtector (KASCADE) experiment~\citep{antoni2003}, which concluded in 2013. The emission spectrum of this scintillator is shown in figure \figref{spect}, with a maximum light output at 434 nm~\citep[][Appendix~A]{kriegleder1992}. This matches well with the maximum PDE of the SiPMs used within the detector.

Each sensor is mounted on a small circuit board developed for this experiment.  The board regulates power to the mounted SiPM, and passes its signal output through two MAR-6SM+ amplifiers; this is an inverting amplifier, but the use of two amplifiers preserves the positive polarity of the signal pulse.  The board also contains fixed attenuation to control the dynamic range, and applies high-pass filtering.

\begin{figure}
    \centering
    \includegraphics[width=\columnwidth]{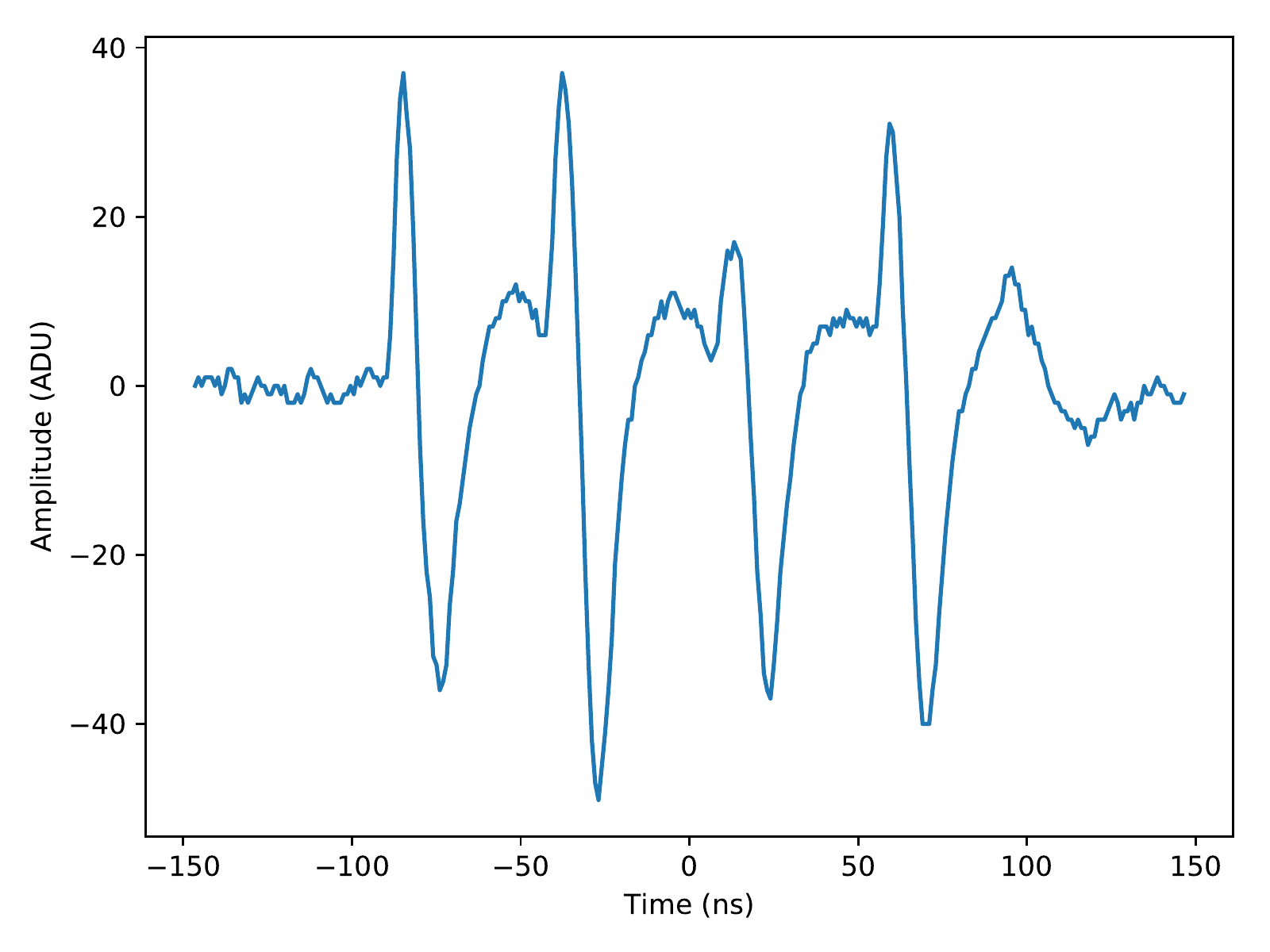}
    \caption{Example raw (unfiltered) trace measured by the SKAPA prototype. Four peaks are visible, each separated by approx.\ 50\,ns, corresponding to the four SiPMs.  The bipolar pulse shape is the result of low-frequency pulse components being blocked by the RFoF link.}
    \label{fig:trace1}
\end{figure}

The output of each SiPM board runs through a separate co-axial cable before the signals are combined.  Each of these delay lines has a different length (0.5m, 10m, 30m and 50m) to ensure that the signals are offset in time by intervals of \mbox{$\sim 50$}~ns.  The resulting output is sent over a single optical fiber, using  RFoF (radio frequency over fiber) links from \emph{Astrotec}.   The pulse polarity, attenuation and high-pass filtering on the SiPM boards are all tuned to maximise the performance and dynamic range of these links. A detection will produce a series of up to 4~pulses at intervals of 50~ns (an example of this is shown in \figref{trace1}).

The fiber runs from the detector to the original equipment hut through about 160~m of single-core fiber. Using an existing link, the signal is then relayed to the Control Building via the fiber patch cabinet, where it is converted back into an electrical signal by another RFoF converter. The signal is then attenuated to match the dynamic range of the ADCs within the data acquisition board, known as the Bedlam Board.

\subsection{Bedlam Board and Control Computer}
\label{sec:backend}

The Bedlam Board~\citep{bray2012} was originally designed for the LUNASKA experiment with the Parkes radio telescope~\citep{bray2014a}. It consists of two analogue-to-digital converters (ADCs) and five field programmable gate arrays (FPGA) as well as an external clock. It can perform 8-bit, 1.024\,GHz digitization on eight input channels, and perform coincidence logic between channels, i.e.\ it will be capable of generating triggers from an array of up to eight particle detectors. Currently, only one of the eight inputs is used for the prototype detector.

The Bedlam Board and its control computer are located within the correlator room of the MRO Control Building. The computer is located directly below the board, with three connections that facilitate the programming and control of the FPGAs as well as data transfer after a trigger.

The analogue signal from the optical fiber is digitized using only one input of the Bedlam Board. A 64-tap matched filter is then applied to the signal to increase its signal-to-noise ratio (SNR). Both the raw and filtered signals are recorded on an internal buffer, and transferred to the control computer upon the filtered signal exceeding a threshold. Both the trigger threshold, and the fraction of buffer returned, are tunable via the control computer.

\subsection{Power supply}
\label{sec:power}

 The prototype detector requires power for the four SiPMs and RFoF connector. This is supplied through 80\,m of cable by a power supply located inside the original equipment hut, which is connected to mains power. The voltage set at the power supply is 34\,V DC, which after a voltage drop of $\sim6$\,V due to cable resistance (depending on the temperature), matches the 25--29\,V operating range of the SiPMs and support boards. The power is low-pass filtered at the gland plate of the original equipment hut and at the input to the detector, minimizing RF emission escaping via the cable.
 
 Following the filtering stage within the detector, the power is passed through a regulator. The optic link support board that powers the fiber optic link is sent 3\,V, while a distribution board sends power to each of the four SiPMs at the full input voltage.
 
 In the future, an electronically adjustable power supply may be used to adjust the input voltage to counteract the effect of both temperature changes on the voltage drop through the power cable, and the temperature-dependence of SiPM behavior. However, in the current set-up, the voltage is fixed at the supply, and hence the voltage at the SiPMs is subject to environmental fluctuations.

\subsection{Radio-frequency interference: mitigation features and testing}

The choice of Murchison Shire for the location of a radio-astronomy observatory is primarily driven by the desire for radio-quietness. However, the traditional photo-sensors used in scintillation detectors are photomultiplier tubes (PMTs), which run at high voltages ($>1$\,kV), and are known to emit short pulses of radio waves (see e.g.\ figure~3 in \citet{apel2010}). The desire to use low voltage devices therefore was the main motivator for using SiPMs as photon detectors. Emissions were further minimized by keeping the particle detector as a purely analogue device, placing it in a metal box, and filtering the voltage on the power supply.

The components of the SKAPA prototype were tested according to the ``RFI Standards for Equipment to be deployed on the MRO'' as defined by CSIRO. Compliance is based on radiated emissions assessment following Military Standard MIL-STD-461F test procedure RE102. The RE102 limits selected by CSIRO are the ‘Navy Mobile and Army’ (NM\&A) limits. Depending on the separation distance to other (i.e.\ non-MWA) instruments at the MRO, additional shielding is required.

The SKAPA prototype was assessed as being in the 1--10\,km range from other MRO instruments (ASKAP and EDGES), requiring 20\,dB of shielding. As shielding was already applied to the prototype, the target emissions threshold of 20dB below RE102 (Navy Mobile \& Army) was used during emissions testing. This assessment was applied to the detector and field power distribution. The limits for the power supply and Bedlam Board/control computer were raised by the shielding effectiveness of the original Equipment Hut and CSIRO Control Building respectively.  All devices were tested against these limits and passed. This facilitated the deployment of the apparatus as shown in \figref{mro_map} during the week of $12^{\rm th}$--$16^{\rm th}$ November 2018.

\section{Preliminary Results}
\label{sec:prelim}

We present a basic characterization of the behavior of the SKAPA prototype, based on data taken from the detector between November 2018 and February 2019.

\subsection{Count rate, dead time and buffer length}
\label{sec:deadtime}

The count rate is the rate at which triggered events (voltage above threshold) are recorded to disk. This is primarily a function of the rate of significant events in the data stream, reduced by the dead time associated with data transfer to the control computer.

To test the dependence of the dead time upon the buffer size setting of the Bedlam Board, the threshold was set so as to saturate the trigger rate. The buffer size was then increased from 256 to 8\,192 samples in powers of 2, with data collected for two minutes in each case. Data were not collected for buffer sizes set below 256 sample as a four-peak event requires approximately 150 samples to be registered; nor were data collected for the buffer size set to 16\,384 as this could not be handled by the control computer.

\begin{figure}
    \centering
    \includegraphics[width=9cm]{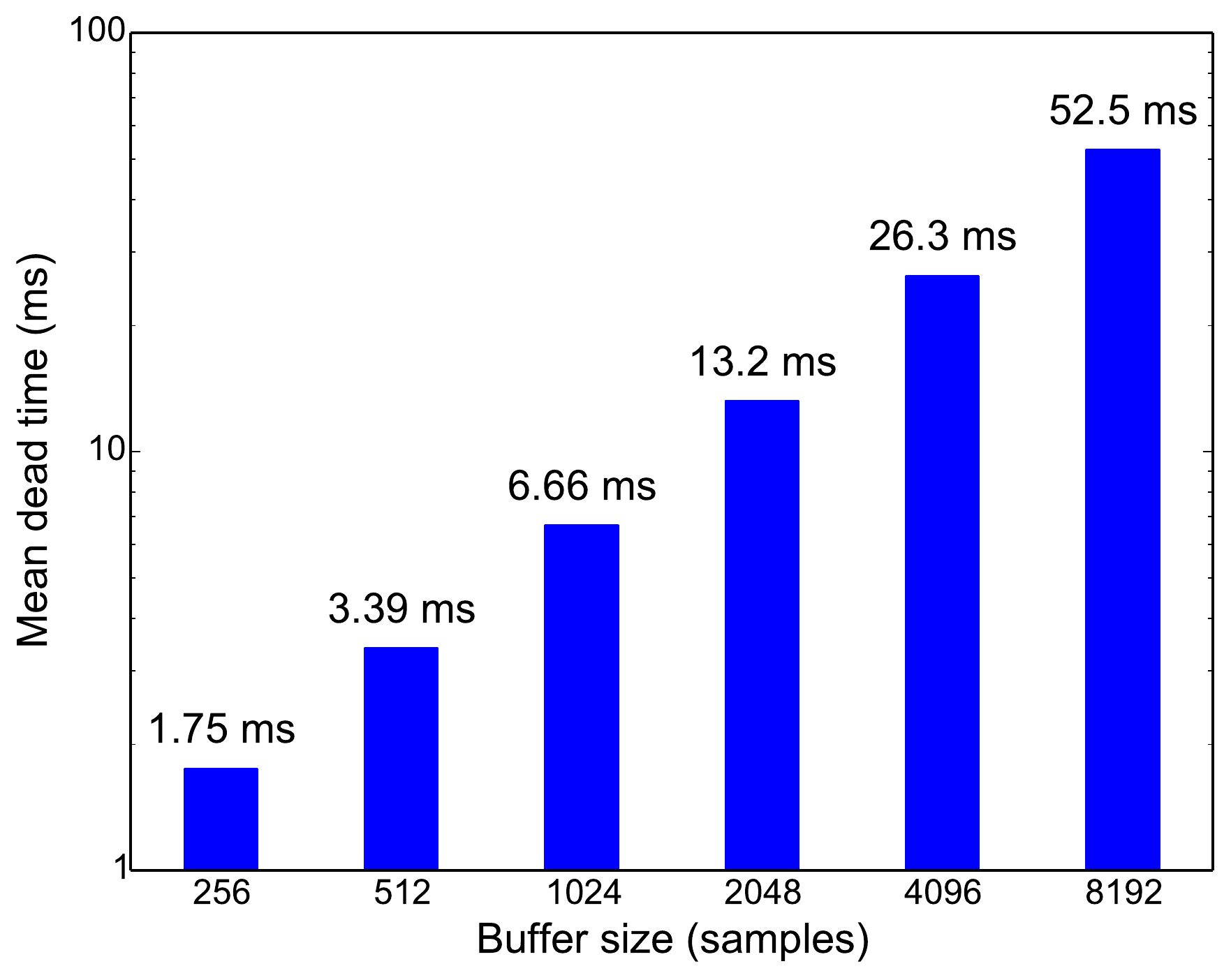}
    \caption{Mean dead time after each trigger event, during which the backend is unable to respond to further triggers, for different buffer-size settings.}
    \label{fig:deadtime}
\end{figure}

The dead times, determined by simply dividing the total number of recorded events by the observation time, are shown in \figref{deadtime}. The dead time is almost linearly proportional to buffer size.

In the following, we use the estimated dead time to correct all observed rates. Thus rates will reflect the true rates in the data, not the rate at which it can be written to disk.

\subsection{Rate vs threshold}
\label{sec:rate_vs_threshold}

To test the dependence of the count rate upon the threshold setting of the Bedlam Board, the buffer size was set to 2\,048, and the threshold increased from 10 to 120\,ADU (analogue-to-digital units) in increments of 10. A threshold equal to the maximum amplitude of 127 was also tested. For each threshold, data were collected for two minutes.

\begin{figure}
    \centering
    \includegraphics[width=\linewidth]{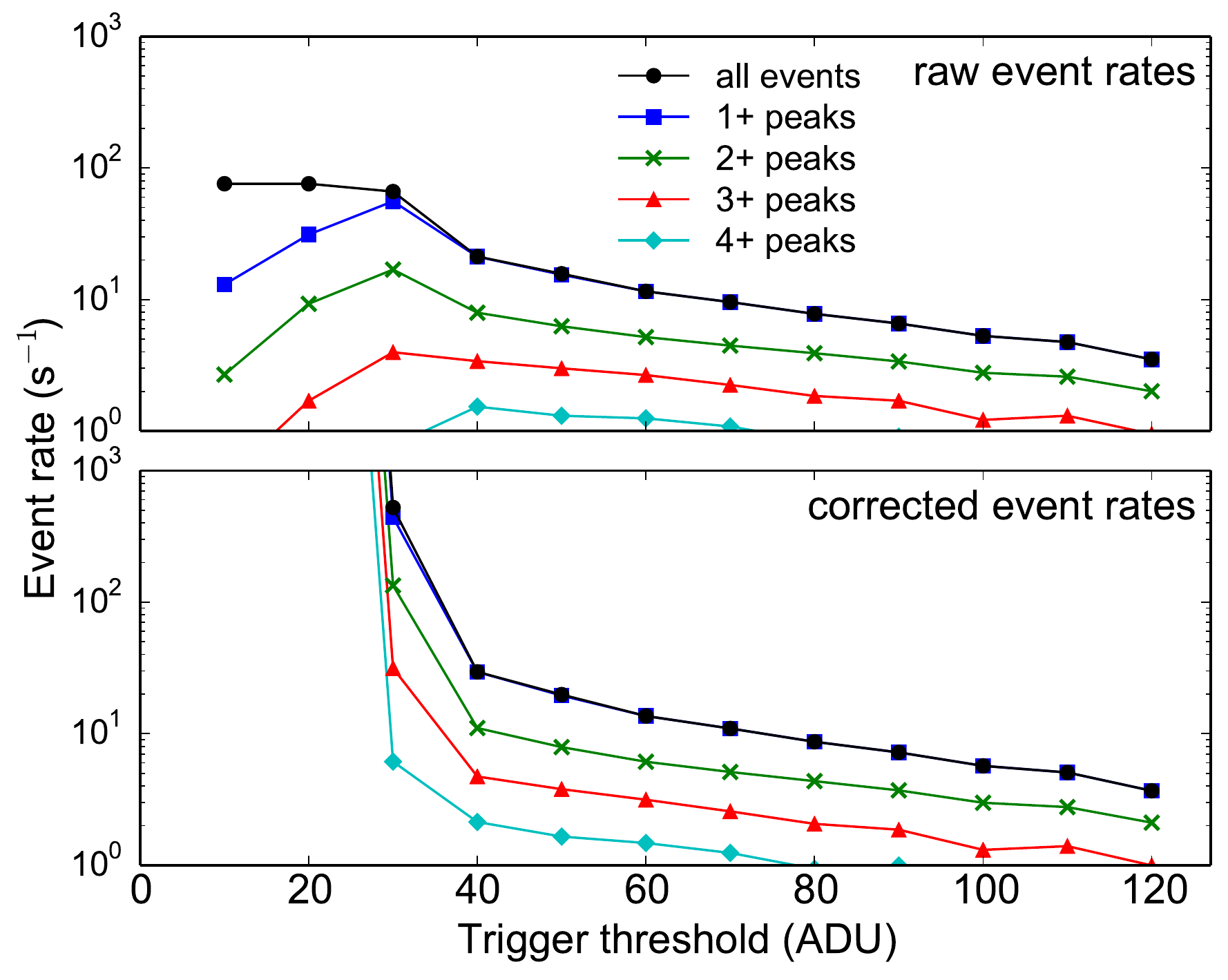}
    \caption{Rates of events classified by the number of peaks in the recovered signal, before (top) and after (bottom) compensating for the loss of triggers due to dead time (see \secref{deadtime}), as a function of the trigger threshold. The extremely high corrected rate at low trigger threshold is due to the deadtime resulting in a very low effective observation time. Data were taken over a two minute period, so the fractional error in a rate $R$ is $(120 R)^{-0.5}$. This is too small to be visible in the figure.}
    \label{fig:rate_vs_threshold}
\end{figure}

The results are represented in \figref{rate_vs_threshold}. The total rate is divided into contributions from events classified by the number of detected peaks (see \appref{peakfinding}). For reference, the accepted experimental rate of muons at sea level is \(\approx1 \,\rm{cm}^{-2} \, \rm{min}^{-1}\)~\citep{tanabashi2018}, i.e.\  \(\approx135\, \rm{sec}^{-1}\) over the $90 \times 90$\,cm$^2$ scintillator surface.

The total count rate shows the expected behavior between thresholds of 30 and 120\,ADU, with a smoothly decreasing count rate with increasing threshold. Below 30\,ADU, the raw event rate saturates, with a poorly defined corrected event rate.

Events with zero detected peaks are considered ``dark count'' events, i.e.\ triggers generated by SiPM cascades initiated by thermal fluctuations. The rate of such events increases rapidly as threshold decreases below 30, and these triggers crowd out single and multi-peak events.

\begin{table}
\centering
\begin{tabular}{c | c c c c}
N [peaks] & 1 & 2 & 3 & 4 \\
\hline
Rate [Hz] & 310 & 102 & 25 & 6\\
\end{tabular}
\caption{Calculated absolute rates of N-peak events using the corrected rates at a threshold of 30\,ADU from \figref{rate_vs_threshold}.} \label{tab:abs_rates}
\end{table}

The approximately constant ratios of $\ge 1$-peak events for thresholds below 30 suggests that the decrease in their total number is purely due to this crowding-out effect. We therefore estimate their true total rate as the corrected rate at a threshold of 30. This is given in \tabref{abs_rates}. The total number of 2+ peak events is comparable to the expected muon rate, indicating that a significant fraction of single-peak events are likely due to radioactive backgrounds or dark-count events (microcells firing without absorbing a photon). A more detailed analysis of the behavior of the detector is given in \secref{montecarlo}.

Given that it is the lowest threshold at which no dark-count events are detected, a threshold of 40 was chosen as the default trigger level for the detector. Note however that this will \emph{not} represent the typical threshold for purposes of a detector array, which will only record events (and hence incur DAQ dead time) upon a multiple coincidence over many detectors.

\subsection{Temperature dependence}
\label{sec:temp-dependence}

The detector is located in an arid climate, with air temperatures ranging from -6\degree to 48\degree C \citep{BoM}. The SKAPA prototype is located in the field, with minimal thermal ballast and no solar shielding. The temperature of the SiPMs themselves will be a function of insolation and the outside air temperate, both with some lag, and a small amount of internal power dissipation. Here, we use the outside air temperature as a proxy for SiPM temperature.

As the temperature increases, so does the thermal energy of electrons in the SiPMs, and microcells in a SiPM can fire without absorbing a photon. The resulting dark count rate for the SiPMs is known to increase exponentially with temperature~\citep{ramilli2008}. Conversely, the gain of SiPMs is known to decrease linearly with increasing temperature~\citep{ramilli2008}, so that photon-induced events will have lower amplitudes. The former effect will result in a greater trigger rate at higher temperatures, the latter effect a lower rate. Both will need to be accounted for when searching for cosmic ray events using an array of detectors.

\begin{figure*}
    \centering
    \includegraphics[width=9cm]{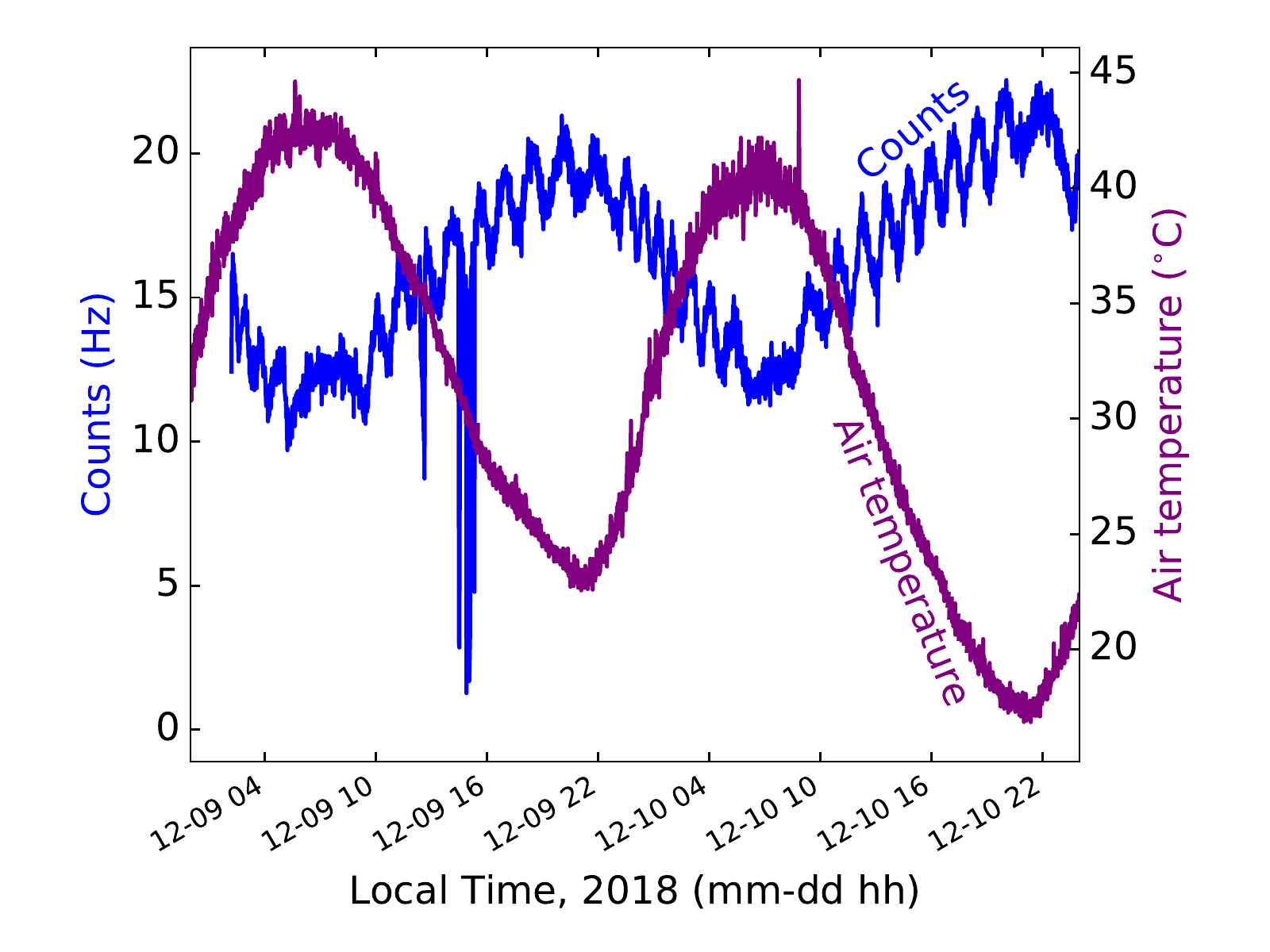}
    \includegraphics[width=9cm]{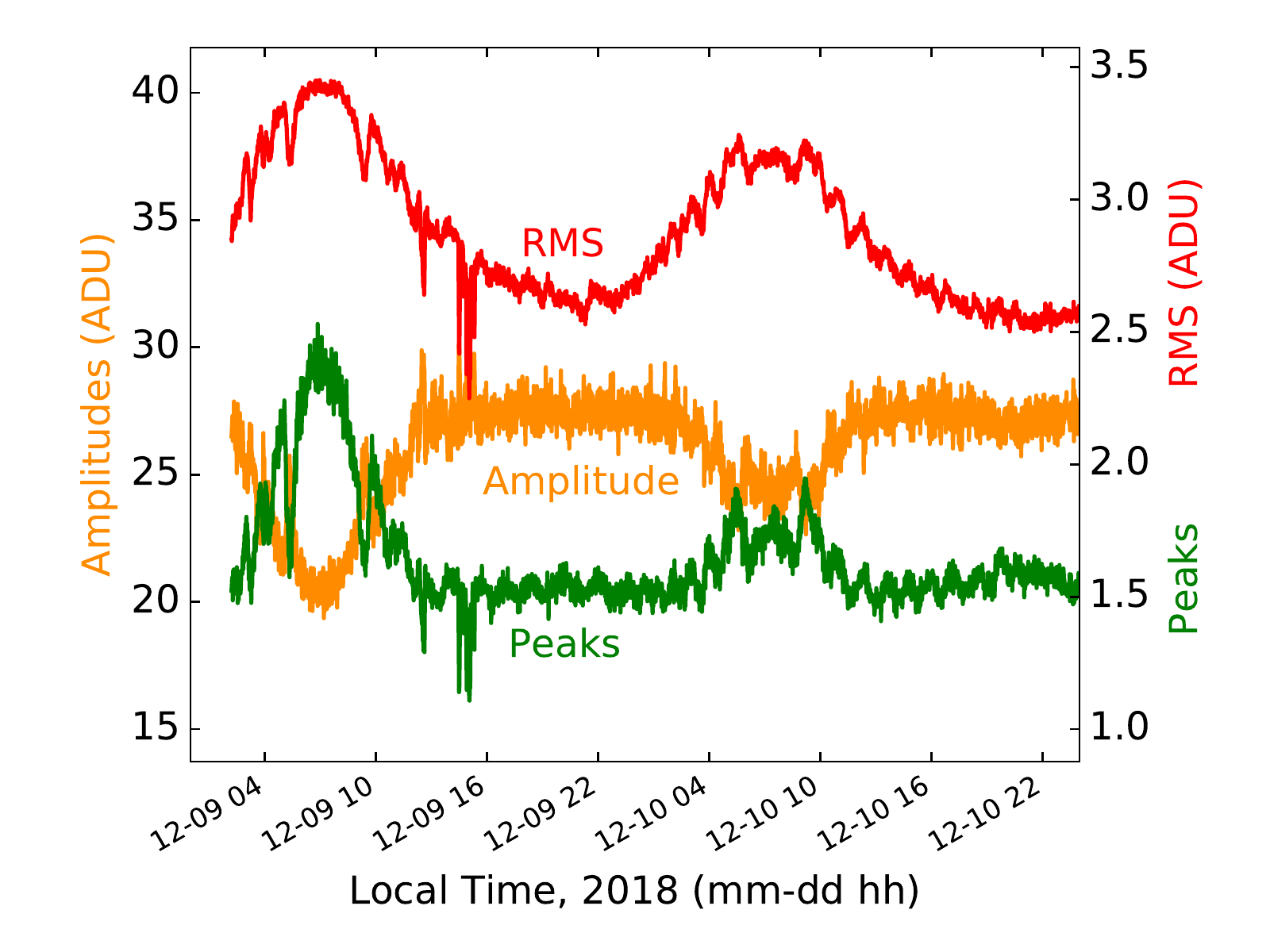}
    \caption{Left: count rate (blue; Hz), and air temperature (purple; $^{\circ}$C) for data taken during the period 9--11 December 2018. Right: average peak amplitude (green; ADU), average number of peaks (orange) and average RMS (red; ADU) for the same period. The dropout near 12-09~16 may be due to a temporary power outage.}
    \label{fig:count_temp_etc}
\end{figure*}

\begin{figure*}
    \centering
    \includegraphics[width=9cm]{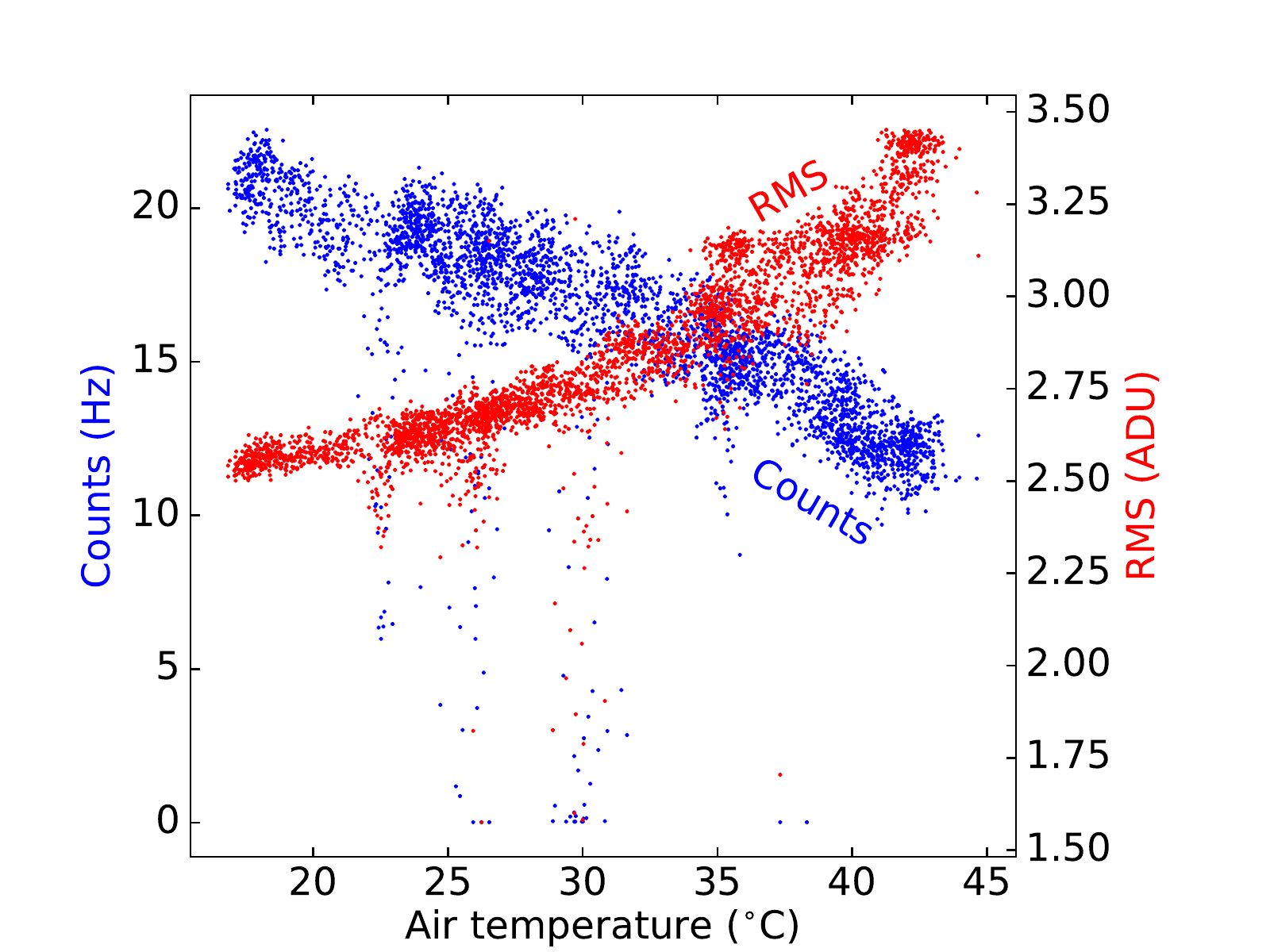}
    \includegraphics[width=9cm]{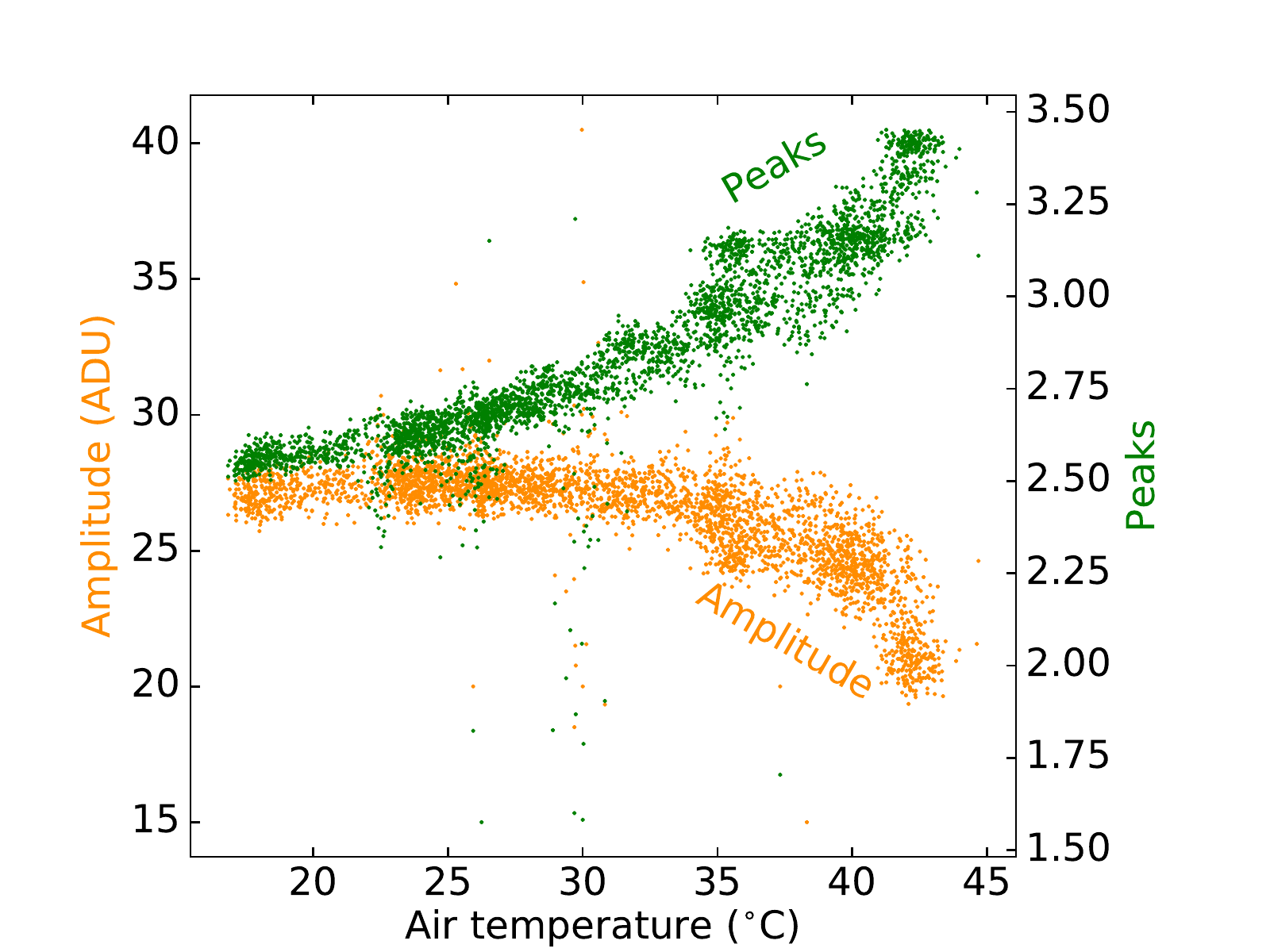}
    \caption{Left: count rate (blue; Hz), and average RMS (red; ADU), as a function of temperature ($^{\circ}$C) for the period 7--11 December 2018. Right: average peak amplitude (orange; $^{\circ}$C), and average number of peaks (green), for the same period.}
    \label{fig:count_vs_temp}
\end{figure*}

In order to observe the expected fluctuations in the count rate, data were collected over a period of 96\,hr from December $7^{\rm th}$--$11^{\rm th}$ 2018. The Bedlam settings were a trigger threshold of 50\,ADU and a buffer size of 2048 samples for this period. The data were analyzed to recover the total trigger rate, the mean number of peaks per trigger, the average peak amplitude, and the rms. This is compared to temperature data from on-site, with \figref{count_temp_etc} plotting all quantities as a function of time, and \figref{count_vs_temp} plotting the count rate, average amplitude, number of peaks, and RMS as a function of temperature.

The primary effect seen in \figref{count_temp_etc} is due to diurnal (day-night) fluctuations in temperature (orange), with a secondary modulation, on timescales of approximately an hour, also being observed. The origin of this secondary oscillation is unknown, but may originate from temperature fluctuations affecting the power supply within the temporary equipment hut.

From \figrefii{count_temp_etc}{count_vs_temp} it is clear that the count rate is anti-correlated with temperature; the correlation coefficient is -0.75, with the remaining scatter being primarily due to the secondary modulation. This is consistent with the gain-reducing effect being dominant, and is further evidence that at standard trigger parameters, the event rate is dominated by real events.

The count rate varying by a factor of $\sim2$ over the temperature range of the sampled data corresponds to an effective change of trigger threshold of $\pm 10$\,ADU (see \figref{rate_vs_threshold}), i.e.\ a $\pm20$\% gain variation.

The increase in the dark-count rate with temperature is evident in \figref{count_vs_temp} as a smoothly increasing RMS value of the frames (red points).  When the temperature exceeds \(\approx33\)\degree C, the average amplitude of peaks begins to decrease, and the average number of peaks in a buffer begins to increase, as seen in \figref{count_vs_temp}. This is consistent with a rapid increase in the number of low-amplitude peaks due to an increasing dark-count rate.

Further investigation revealed that the change in the total count rate is dominated by a change in the rate of single-peak events, with excess multi-peak events only being significant above \(\approx35\)\degree C.

In conclusion, we consider normal operating conditions for the detector to be at thresholds of 40\,ADU and above, and at temperatures below \(\approx35\)\degree C. In this regime, we expect that 2+ peak events are predominantly due to atmospheric muons, and that the detection rate is stable. For higher temperatures, active adjustment of the trigger threshold, or the SKAPA supply voltage, will be required to stabilize the detection efficiency.

\section{Simulations}
\label{sec:montecarlo}

The time lags between the output of the four SiPMs inside the SKAPA prototype allow the combinations of SiPMs which have fired in each event to be identified. For instance, a 3-peak event, with lags between the peaks of 50 \& 100\,ns, corresponds to SiPMs 1, 2, and 4. The likelihood of any given SiPM firing will clearly be correlated with the event geometry. Analyzing the rates of each event class allows a far more detailed analysis of the detector performance than simply the N-peak coincidence rates analyzed in \secref{prelim}.

In this Section, we develop a toy model of the detector response to compare to these rates. The success of this model is then used to extrapolate to the performance of the planned detector array.

\subsection{Description of the model}
 \label{sec:model_description}

The detector model considers the flux of through-going muons at sea level described by \citet{tanabashi2018}. Thus, a total muon rate of \(\approx135\, \rm{sec}^{-1}\) was assumed, with zenith-angle distribution $p(\theta_z) \propto \sin \theta_z \cos^3\theta_z$ for a flat detector.  The average muon energy is 4\,GeV. At this energy, 99.8\% of energy loss is due to ionization with minimal contribution from Bremsstrahlung, pair-production and photonuclear radiation processes.

All muons were therefore treated as minimum-ionizing particles, with energy loss 2.37\,MeV\,cm$^{-1}$ \citep{MuonStoppingPower}, producing $n_\gamma=23,700$ photons\,cm$^{-1}$ in the scintillator \citep{tanabashi2018}. Ignoring edge effects, the total number of photons, $N_\gamma$, produced by a through-going muon will therefore be:
\begin{eqnarray}
N_{\gamma} & = & \frac{h_{\rm s} n_\gamma}{ \cos \theta_z}
\end{eqnarray}
for a scintillator block height $h_{\rm s}=3$\,cm.

The photons from each muon were assumed to emanate from the impact point, be distributed uniformly in solid angle, and propagate ignoring absorption or scattering.  The typically many reflections that a photon will undergo from the upper and lower scintillator surfaces implies that any reflection coefficient less than unity makes escape likely. Therefore, all photons produced with incident angle less than the angle of total internal reflection ($\theta_{\rm tir}=38.7^{\circ}$ given $n=1.605$ for photons of (assumed monochromatic) wavelength $\lambda=430$\,nm) were discarded, and the rest assumed to propagate only in the horizontal plane. This reduced the total number of photons to the fraction $f=0.78$, i.e.\ 22\% were assumed to escape.

The expected number of photons $\left<N_{\rm in}\right>$ incident on each SiPM was then calculated using:
\begin{eqnarray}
\left< N_{\rm in} \right> & = & f N_\gamma \frac{w_{\rm SiPM}}{2 \pi R} \frac{h_{\rm SiPM}}{h_{\rm s}} \hat{r} \cdot \hat{n}_s \nonumber \\
& = & f \frac{n_\gamma}{ \cos \theta_z} \frac{1}{2 \pi R} \hat{r} \cdot \mathbf{A}_{\rm SiPM}
\end{eqnarray}
for an event located a distance $R$ (unit vector $\hat{r}$) from a SiPM with normal vector $\hat{n}$, width $w_{\rm SiPM}$, height $h_{\rm SiPM}$, and surface $\mathbf{A}_{\rm SiPM} = w_{\rm SiPM}\, h_{\rm SiPM} \,   \hat{n}$. Photons reflected at the sides of the scintillator were ignored. This formula gives $\left< N_{\rm in} \right> \approx 120$ incident photons for a vertical muon in the centre of the detector.

The probability of an incident photon firing a SiPM microcell is the photon detection efficiency (PDE). We ignore its angular and wavelength (\figref{spect}) dependencies, and take the value for normal incidence at $430$\,nm, i.e.\ PDE=0.4. Once one microcell fires, the probability of a nearby microcell firing (crosstalk) is $p_\chi(V)$, where $V$ is the over-voltage. For the overvoltage used here, $p_\chi=0.1$. The total expected number $\left<F\right>$ of fired microcells for each event is therefore:
\begin{eqnarray}
\left<F\right> & = & {\rm PDE} \left(1+p_\chi(V)\right) \left< N_{\rm in} \right>. \label{eqn:fired_microcells}
\end{eqnarray}
Given this expectation ($\left<F\right> \approx 65$ for a centrally located vertical muon), the actual number of fired microcells was sampled using a Poisson distribution, i.e.\ ignoring the mutual dependence of the cross-talk effect.

Rather than model the analogue signal chain, we aim to determine a linear scaling relationship between the characteristic number of fired microcells from \eqnref{fired_microcells} and the detector threshold in ADU, in order to estimate the behaviour of a detector array where a lower individual-detector threshold will be used. The unknown constant in this relationship formally has units of ADU/microcell, i.e.\ it is the expected amplitude of the pulse peak after digitization induced by the firing of one microcell.

\subsection{Trigger simulation and event classification}

\begin{table}
\centering
\begin{tabular}{c c c}
N$_{\rm peak}$ & Pattern & SiPMs \\
\hline
4 & 50-50-50 & 1,2,3,4 \\
\hline
3 & 50-100 & 1,2,4 \\
  & 100-50 & 1,3,4 \\
  & 50-50 & 1,2,3; 2,3,4 \\
\hline
2 & 50 & 1,2; 2,3; 3,4 \\
  & 100 & 1,3; 2,4 \\
  & 150 & 1,4 \\
\hline
1 & N/A & 1; 2; 3; 4\\
\end{tabular}
\caption{List of event classes identifiable by the peak-finding algorithm, giving the number of peaks $N_{\rm peak}$, the pattern of observed time separations between peaks (ns), and the combinations of SiPM detections this corresponds to.} \label{tab:event_categories}
\end{table}

For each simulated event, a trigger was assumed to occur whenever the signal in any SiPM was modeled as equaling or exceeding some number of microcells, $F_{\rm th}$. The event classification (see \tabref{event_categories}) however is determined by the peak-finding algorithm (\appref{peakfinding}), which uses a second, lower threshold, corresponding to approximately half of the primary 50\,ADU threshold. This was therefore set to $0.5 F_{\rm th}$, with SiPMs detecting more than this many photons being used for the event classification.

\begin{figure}
    \centering
    \includegraphics[width=\columnwidth]{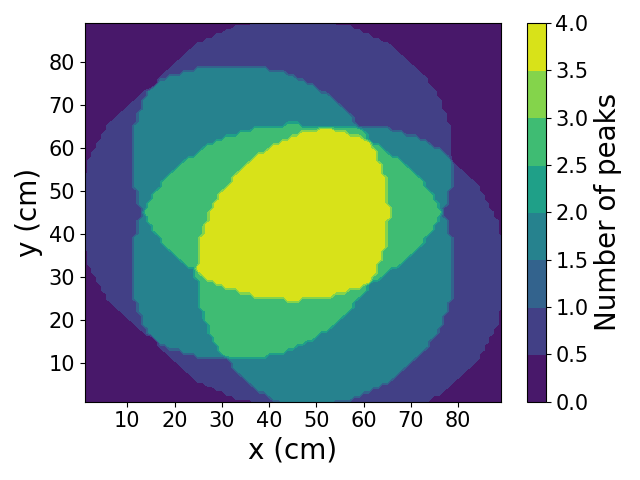}
    \caption{Plot of the horizontal plane of the scintillator. The color scale illustrates which modeled vertical muon positions produce above-threshold triggers on 0, 1, 2, 3, and 4 of the SiPMs when Poisson fluctuations are removed. The assumed threshold is 25 fired microcells. Note that SiPM~1 (right) is physically offset from the mid-point of the scintillator.} \label{fig:2d_deterministic}
\end{figure}

Removing Poisson fluctuations produces a 1--1 relationship between event location and its classification, with that relationship depending on $F_{\rm th}$ and incident muon angle. This is illustrated in \figref{2d_deterministic}, in the case of vertical muons, and a threshold of $F_{\rm th}=25$ fired microcells. The asymmetry caused by the offset of SiPM~1 is evident --- the region corresponding to three-fold events on SiPMs 1,3,4 (i.e.\ not 2) is very small. Two-peak events from SiPMs 1,3 and 2,4 are impossible without random fluctuations. The corners of the scintillator block are also not very sensitive.

\subsection{Results, and comparison to data}
\label{sec:mc_results}

\begin{figure*}
    \centering
    \includegraphics[width=10cm]{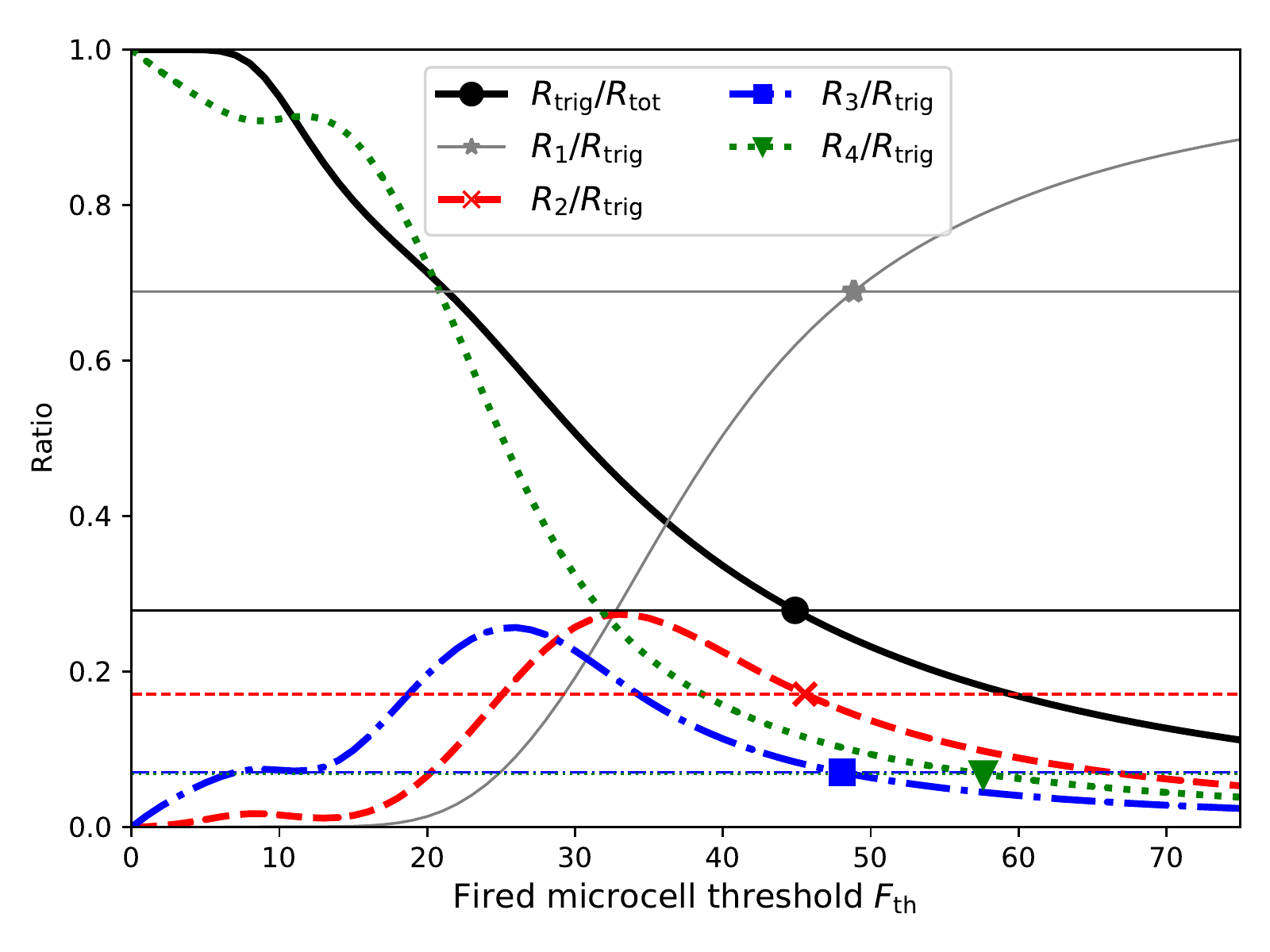} \\
    \includegraphics[width=10cm]{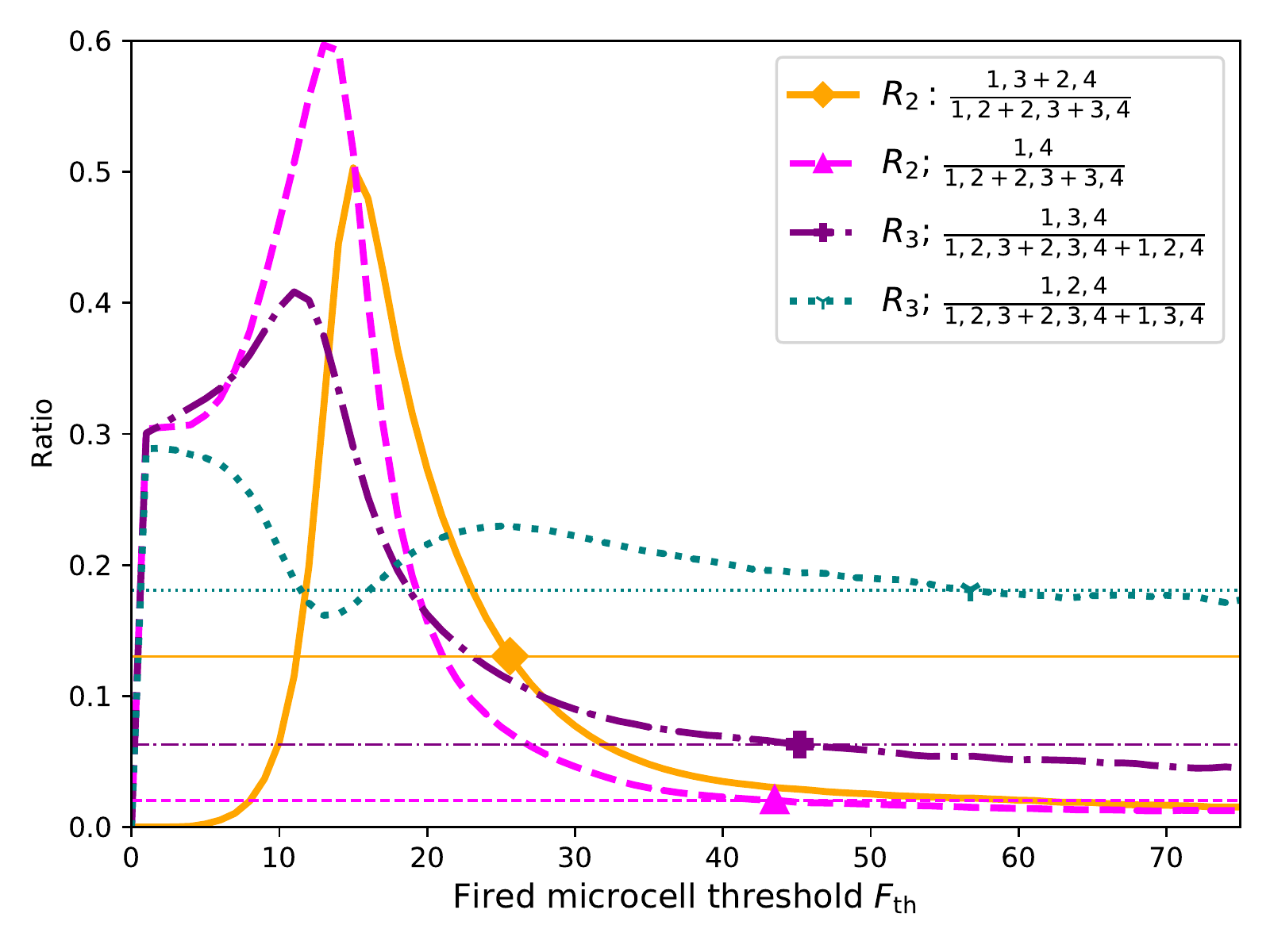}
    \caption{Comparisons between measured (horizontal) and simulated (varying) ratios of events of different classifications, as a function of the number of simulated fired microcells, $F_{\rm th}$, required for a primary trigger. Top: the total triggered rate $R_{\rm trig}$ is compared to the total muon rate $R_{\rm tot}$, and the rates $R_{\rm N}$ for N-peak events are shown compared to $R_{\rm trig}$. Bottom: for 2-peak and 3-peak events, ratios between different event classes that can be identified in data are also given (see \tabref{event_categories}). In each figure, the markers indicate the number of fired microcells at which the simulated rates agree with the measurement, assessed independently for each event class. Note that the measured rate of $R_3$ and $R_4$ is almost identical.}
    \label{fig:mc_results}
\end{figure*}

The simulated ratios of all different event classifications that can be identified in data (see Table~\ref{tab:event_categories}) are shown as a function of trigger threshold in units of fired microcells, $F_{\rm th}$, in \figref{mc_results}. These are compared to event rates from data taken over a 16\,hr period, using a 50\,ADU threshold, and a buffer length of 8192 samples. The corresponding rates of each event category, corrected for dead-time, are shown as horizontal lines. Intersection points between measurements and model predictions are also indicated. Where multiple intersections occur, only the most likely value is indicated. In an ideal case, all intersections will occur for the same, true, value of $F_{\rm th}$.

From \figref{mc_results}, the modeled total event rate (black), and rates of 1--3 peak events (gray, red, and blue), intersect the observed event rates when the threshold corresponds to $F_{\rm th}=46\pm2$ fired microcells. Also in close agreement are the relative fractions of 2-peak events between SiPMs 1 and 4 (pink), and of three-peak events between SiPMs 1,3, and 4 (purple).

The disagreement between measured and expected rates of 3-peak events involving SiPMs 1, 2, and 4 (teal) at $F_{\rm th}=46$ is very small, with the intersection point between measurement and expectation being so far offset due to the shallow slope of the expectation curve.

The two points showing significant disagreement however are the rates of 4-peak events (green: the measured rate is below expectation for $F_{\rm th}=46$), and the fraction of 2-peak events involving SiPMs 1 and 3 or 2 and 4 (orange: the measured rate is far above expectation at $F_{\rm th}=46$).

The reason for these discrepancies is that \eqnref{fired_microcells} likely over-estimates the expected number of fired microcells. For instance, it ignores absorption/scattering of photons within the scintillator, and assumes 100\% transmission of photons through the optical gel at the scintillator--SiPM interface. To first order, any effect leading to a decrease in $\left< F \right>$ would merely result in a shift in the values plotted on the $x$-axis in \figref{mc_results}. However, as the mean decreases, the importance of Poisson fluctuations about the mean will increase.

This effect most obviously explains the observed excess in 2-peak events with SiPMs 1,3 and 2,4, which as discussed in relation to \figref{2d_deterministic}, only occur due to Poisson fluctuations. It also explains the under-occurrence of four-peak events. An under-fluctuation in the actual number of fired microcells on any of four SiPMs can result in a 4-peak event registering as a 3-peak event, whereas there are no 5-peak events providing a compensating increase in the 4-peak rate.

A short investigation has revealed that artificially increasing the Poisson fluctuations by a factor of $\sqrt{5}$ (i.e.\ decreasing $\left< F \right>$ in \eqnref{fired_microcells} by a factor of 5) corrects this disagreement. However, we do not pursue this further, since the goal is not to perfectly model the behavior of this prototype, but to model it sufficiently well to predict the behavior of an array, as discussed in the following Section.

We therefore conclude that a threshold of $F_{\rm th}=46 \pm 2$ microcells corresponds to a trigger threshold of 50\,ADU, i.e.\ there is a scaling constant of $1.09 \pm 0.05$\,ADU per fired microcell.

\section{Estimation of array performance}
\label{sec:performance}

The aim of this project is to deploy an array of several SKAPA detectors at the MRO, and use multiple coincidences to form a trigger on extensive air showers. Such an array will use a two-stage trigger: a simple threshold trigger on each individual detector, similar to that used by the SKAPA prototype analyzed here; and a secondary trigger, requiring N first-stage triggers to occur within a time window W. With the information from the prototype detector, we can now estimate the sensitivity of this set-up.

\subsection{Modified stage 1 trigger}

The SKAPA prototype's use of delay lines to each SiPM allowed the detailed evaluation of detector performance given in \secref{montecarlo}. However, by spreading multi-PMT signals out in time, it reduced the sensitivity to through-going muons. Removing the delay lines will stack all SiPM signals on top of each other, without affecting the rate of false events. It will also remove attenuation in the lines.

\begin{figure}
    \centering
    \includegraphics[width=\columnwidth]{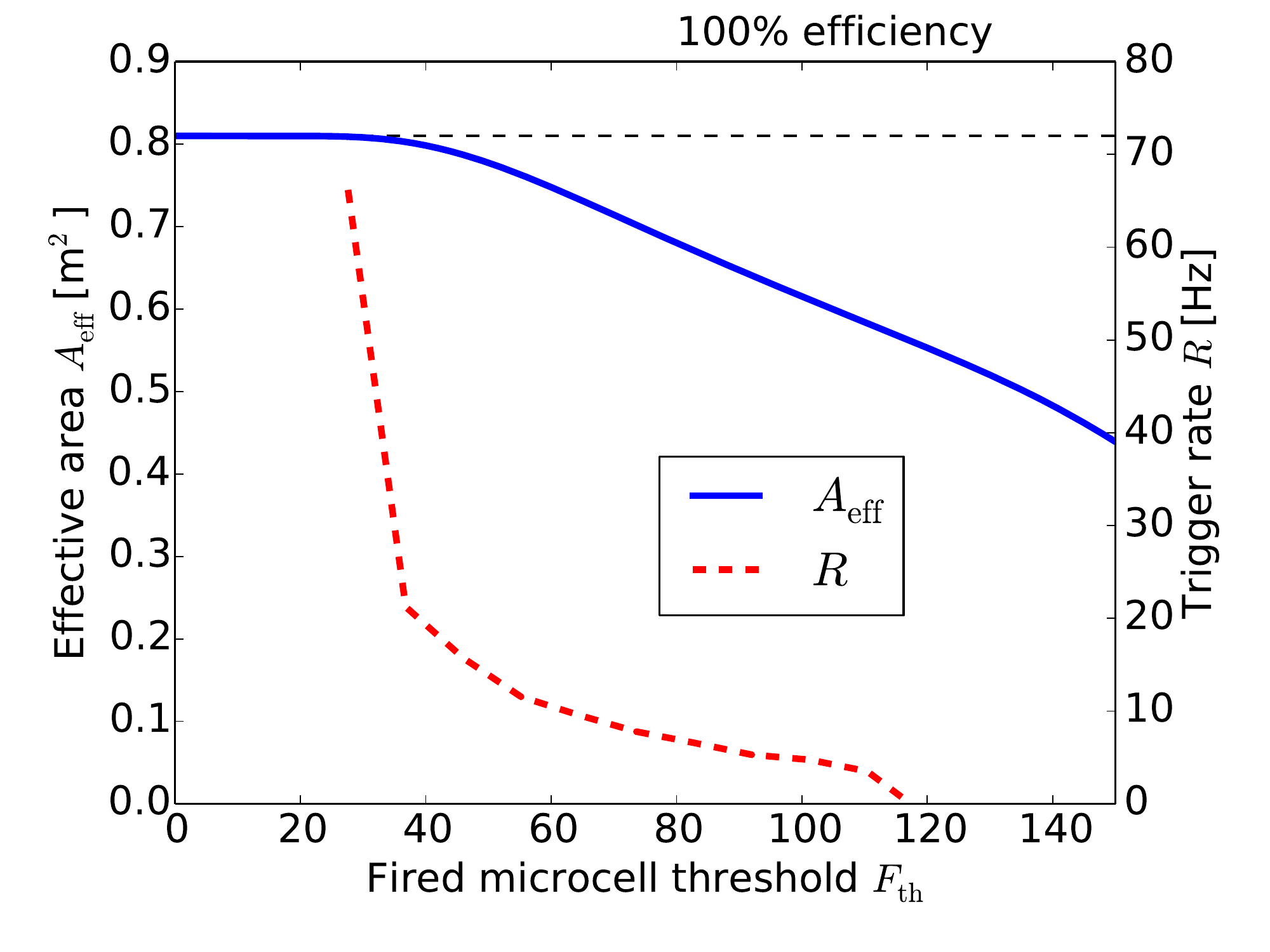}
    \caption{Effective area $A_{\rm eff}$ of a single SKAPA particle detector to vertical muons as a function of the threshold number of fired microcells $F_{\rm th}$, in a possible set-up when no delay lines are used in the final design. This is compared to the expected individual-detector trigger rate $R$ as a function of $F_{\rm th}$.}
    \label{fig:summed_output}
\end{figure}

Modifying the simulation of \secref{montecarlo} to this scenario produces the total trigger efficiencies given in \figref{summed_output}. These are shown as effective areas $A_{\rm eff}$, by multiplying by the physical area of (90\,cm)$^2$. The gain from summing all SiPMs is evident when comparing to the total detection rate (black line) in \figref{mc_results}: where previously only 50\% of events triggered at $F_{\rm th}=30$, removing the delay lines results in 100\% efficiency at this threshold, and more than 50\% efficiency even at $F_{\rm th}=100$.

\subsection{Stage 2 trigger}

The simplest kind of stage 2 trigger would use a sliding window of time width $W$, and require $N$ SKAPA detectors to trigger within this window. Obviously, more physical constraints could be placed on this trigger, e.g.\ requiring detection times to be consistent with the shape of the particle wavefront. Here, we first explore whether or not a high-efficiency trigger would be possible with simple real-time logic before considering more-complicated options.

Considering an array of SKAPA detectors located in the MWA core, we estimate a maximum array baseline of 400\,m, with a light travel time of $w=1.3$\,$\mu$s. We therefore consider a total time window of $w=2.6$\,$\mu$s. If $N$ detectors out of a total of $M$ (nominally, $M=8$) are required to trigger within this window, then the rate of stage 2 triggers, $R_2$, can be calculated as:
\begin{eqnarray}
R_{2} & = & R_1 (^M_N) (R_1 w)^{N-1}. \label{eqn:combinatorics}
\end{eqnarray}
Setting $R_{2}$ to minimize the dead-time  (e.g.\ to 5\%) would allow a trigger rate of up to $1$\,Hz even for a maximal $8,192$-sample buffer. This would generate far too much false radio data. However, a false alarm rate comparable to the expected cosmic ray rate (1\,hr) must, by definition, be acceptable, or otherwise real events will necessarily be missed.

\begin{table}
    \centering
    \begin{tabular}{c c c c}
        N$_{\rm trig}$ & $R_1^{\rm max}$ [Hz] & $F_{\rm th}$ & $A_{\rm eff}$ [m$^2$]\\
        \hline
        2	&	2 	& 110--120 & 0.58--0.54 \\
	    3	&	90 	& $\sim 28$ & 0.81 \\
	    4	&	690 & $\ll 28$    & 0.81 \\
    \end{tabular}
    \caption{Number of stage 1 triggers $N_{\rm trig}$ in the array required for a stage $2$ trigger; maximum stage 1 trigger rate $R_{1}^{\rm max}$ corresponding to a stage 2 rate $R_2$ of once per hour; the trigger level $F_{\rm th}$ producing $R_1^{\rm max}$; and the corresponding effective area, $A_{\rm eff}$, of an individual SKAPA detector to vertical muons.} \label{tab:req_rates}
\end{table}

Requiring $N=$2, 3, or 4 of $M=8$ detectors to trigger within the $2.6$\,$\mu$s window allows the maximum individual rates $R_{1}^{\rm max}$ shown \tabref{req_rates}. Requiring only $N=2$ requires a very low $R_1$, which can be estimated by extrapolating \figref{rate_vs_threshold} to correspond to 120--130\,ADU, i.e.\ $F_{\rm th}\approx$110--120, and a reduced effective area of $A_{\rm eff}\approx 0.65$\,m$^2$. Conversely, requiring $N=4$ allows very high individual rates, which cannot be measured by \figref{rate_vs_threshold}. However, setting $N=3$ allows $R_1=28$\,Hz, where the SKAPA detectors are estimated to already be at 100\% efficiency from \figref{summed_output}. We therefore estimate that it will be possible to use an array of 8 SKAPA detectors while requiring either a 2-fold or 3-fold detector coincidence. Whichever of these settings is superior --- and the optimum detector layout --- will require detailed simulations using CORSIKA \citep{heck1998,heck2017} and CoREAS \citep{huege2013b}, and is beyond the scope of this work.

\section{Conclusion}
\label{sec:conclusion}

The deployment of the prototype Square Kilometre Array Particle Array (SKAPA) detector at the Murchison Radio-astronomy Observatory (MRO) has demonstrated the feasibility of cosmic-ray detection in a remote radio-quiet environment. As of writing, the detector is still functioning, despite enduring air temperature fluctuations from 1--45\,$^{\circ}$C, periods of daily rainfall up to 47\,mm, and power outages associated with maintenance operations. While this nine-month period is significantly less than a targeted ten-year operational period for any detector array deployed at the site, it is a strong indication of the viability of this experiment.

The sensitivity of this detector has been calibrated against temperature fluctuations in the range 18--45\,$^{\circ}$C, where a decreasing overvoltage with increased temperature leads to a $\pm20$\% change in the amplitude of signals from through-going muons. This could be adequately compensated-for either through varying the supply voltage, or altering the stage 1 trigger thresholds. 

The use of delay cables to separate the signals from each of the four SiPMs has allowed the development of a detector model through comparison of predicted and measured event rates for different event classes. Using this model, we estimate that a satisfactory trigger for MWA data could be constructed by either requiring a two- or three-fold coincidence over multiple SKAPA detectors within a time window of $2.6\,\mu$s.

The next stage of SKAPA development would remove the delay cables to increase detector sensitivity, and use fiber and power cables with improved shielding to ensure a longer experimental lifetime. The current target is eight such detectors to cover the core of the MWA, which is the most that could be handled with the current data acquisition system (Bedlam). It is important to note that the dynamic range of the RFoF connection will limit the ability of an array of these detectors to independently reconstruct the particle core. Since the goal however is to trigger radio data-taking, we do not see this as an important constraint for the current system.

We have not yet performed a simulation of the total sensitivity of the planned detector array to extensive air showers visible to the MWA. However, the scintillator blocks are identical in size to those used by the Lofar Radboud air shower array \citep[LORA; ][]{thoudam2011}, and for our proposed three-fold trigger, we expect the required threshold on each SKAPA detector to correspond to 100\% efficiency to through-going muons. Since LORA also requires a three-fold coincidence between detectors within sub-arrays, our resulting trigger is therefore expected to be no less sensitive than that used by the cosmic-ray detection pipeline at LOFAR \citep{schellart2013}. We therefore expect to be able to reliably trigger on cosmic ray events incident above the array with an energy threshold in the 10$^{16}$--10$^{17}$\,eV range.

\section*{Acknowledgments}

We thank ASTRON for the use of the RFoF links in the detector, and the KASCADE-Grande Collaboration for the plastic scintillator used in construction. This scientific work makes use of the Murchison Radio-astronomy Observatory (MRO), operated by CSIRO. We acknowledge the Wajarri Yamatji people as the traditional owners of the Observatory site. Support for the operation of the MWA is provided by the Australian Government (NCRIS), under a contract to Curtin University administered by Astronomy Australia Limited. We would like to thank CSIRO Astronomy and Space Sciences and the MWA Collaboration for the use of facilities at the MRO, and for assistance with deployment operations and pre-deployment testing of the detector. 
Parts of this research were conducted by the Australian Research Council Centre of Excellence for All-sky Astrophysics (CAASTRO), through project number CE1101020; and were supported through Australian Research Council Discovery Project DP200102643. A.W.\ would like to thank Innovation Central Perth, a collaboration of Cisco, Curtin University, Woodside and CSIRO's Data61, for their PhD scholarship. J.B. acknowledges the support of the European Research Council Starting Grant \emph{Advancing astroparticle physics with next-generation radio-astronomy instruments}, and the UK Science and Technology Facilities Council UK-Australia Visit Programme \emph{Processing large data volumes for radio observations and computational simulations of cosmic-ray air showers}. K.M.L.G.\ and R.T.\ acknowledge support from the Ogden Trust. J.B.\ and R.S.\ acknowledge the support of the University of Manchester Investing in Success grant \emph{Augmented cosmic-ray detection at the Murchison Radio-astronomy Observatory}, and the UK Science and Technology Facilities Council Early Career Impact Fund \emph{Particle-detector development for the Square Kilometre Array}.
This research has made use of NASA's Astrophysics Data System Bibliographic Services.

\appendix

\section{Peak-finding algorithm}
\label{app:peakfinding}

Traces recorded by the Bedlam Board are expected to have between zero and four pulses from the four SiPMs in the detector. The time offsets between multi-peak events will also encode information on the geometry of the event (see \secref{montecarlo}).

In order to identify these in digital data, a simple peak-finding algorithm was developed. The algorithm uses the Python function scipy.signal.find\_peaks \citep{SciPy,SciPy2019}, which compares the values of an array to their neighbors to locate local maxima, and can apply additional criteria such as a minimum height to take a subset of the peaks.

The bandwidth limitation of the optical fiber (\secref{sigpath}) turns the intrinsically monopolar SiPM readout into a bipolar pulse. By-eye analysis of the pulses showed that the negative amplitude of the peak in digital data tended to be stronger than the positive peak. Therefore, a peak was defined to be the most negative value within 20~samples, to avoid confusion between adjacent SiPM signals. The threshold for defining a pulse was set at a reading of -9\,ADU on the raw (unfiltered) data, which was sufficient to exclude most dark count events. This corresponded to approximately 25\,ADU on the filtered data.

\begin{figure}
    \centering
    \includegraphics[width=8cm]{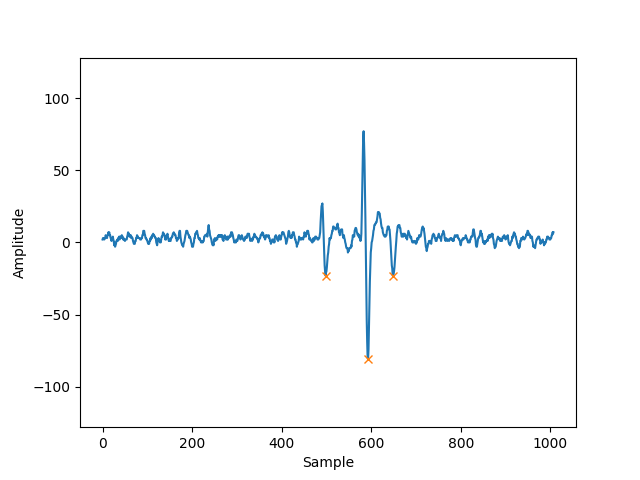}
    \caption{Example event (unfiltered data) from the SKAPA prototype, showing the locations of three peaks found with the peak-finding algorithm (yellow crosses).}
    \label{fig:3_peak_event}
\end{figure}

\figref{3_peak_event} displays an event with three peaks identified using this definition. This algorithm was used to characterize the number, magnitude, and time of SiPM signals in Bedlam-triggered data.

\section*{References}

\bibliographystyle{plainnat}
\newcommand{\urlprefix}{}
\bibliography{journals_full.bib,main.bib}

\end{document}